\documentstyle[12pt,aaspp4]{article}

\newcommand\etal { {\it et al$.$} }
\newcommand\kms  { km~s$^{-1}$ }

\begin{document}

\title{A Photometric and Kinematic Study of AWM 7 \footnote{
Observations reported in this paper were obtained at the
MDM Observatory, a facility jointly operated by the University of Michigan,
Dartmouth College, and the Massachusetts Institute of Technology; 
at the Multiple Mirror Telescope Observatory, a facility operated jointly
by the University of Arizona and the Smithsonian Institution; and at the Whipple
Observatory, a facility operated jointly by the Smithsonian Astrophysical
Observatory and Harvard University.} }

\author{Daniel M. Koranyi and Margaret J. Geller}
\affil{Harvard-Smithsonian Center for Astrophysics}
\authoraddr{60 Garden St., Cambridge, MA 02138}
\authoremail{dkoranyi@cfa.harvard.edu, mgeller@cfa.harvard.edu}
 
\author{Joseph J. Mohr}
\affil{Dept. of Physics, University of Michigan}
\authoraddr{500 E. University, Ann Arbor, MI 48109}
\authoremail{jjmohr@umich.edu}

\author{Gary Wegner}
\affil{Dept. of Physics \& Astronomy, Dartmouth College}
\authoraddr{6127 Wilder Lab., Hanover NH, 03755}
\authoremail{wegner@kayz.dartmouth.edu}

\begin{abstract}

We have measured redshifts and Kron-Cousins $R$-band magnitudes for a sample of
galaxies in the poor cluster AWM 7.  We have measured redshifts for 172
galaxies; 106 of these are cluster members.
We determine the luminosity function from a photometric survey of the
central $1.2 \, h^{-1} \times 1.2 \, h^{-1}$ Mpc.  The LF has a bump at the
bright end and a faint-end slope of $\alpha = -1.37\pm0.16$, populated almost
exclusively by absorption-line galaxies.
The cluster velocity dispersion is lower in the
core ($\sim$530 km~s$^{-1}$) than at the outskirts ($\sim$680 km~s$^{-1}$),
consistent with the cooling flow seen in the X-ray. 
The cold core extends $\sim$150$\, h^{-1}$ kpc from the cluster
center.  The Kron-Cousins $R$-band mass-to-light ratio of the system is
$650\pm 170 \, h \, M_\odot/L_\odot$, substantially lower than previous optical
determinations, but consistent with most previous X-ray determinations.
We adopt $H_0 = 100 \, h$ km~${\rm s}^{-1}$~${\rm Mpc}^{-1}$ throughout this
paper; at the mean cluster redshift,
($5247\pm 76$ km~s$^{-1}$), 1~$h^{-1}$ Mpc subtends 65$\farcm$5.

\end{abstract}


\section{Introduction}

The AWM and MKW clusters were selected on the basis of the appearance of
the central cD galaxy
(Morgan \etal\ 1975, Albert \etal\ 1977).  There is controversy about
the history of these systems and the presumably related formation of the
central galaxy.  Ostriker \& Tremaine (1975) suggest that a
cD galaxy grows by accreting other galaxies through dynamical friction
and tidal stripping; Merritt (1985) suggests that galaxies merge
during the original cluster collapse.  Recent $N$-body simulations
by Bode \etal\ (1994) and Dubinski (1997) produce giant central elliptical
galaxies through hierarchical merging.  Fabian (1994) proposes that
enhanced star formation at the cluster center resulting from a cooling flow may
further enlarge these merged galaxies.

The AWM/MKW clusters span a broad range
of velocity dispersions ($\sigma \sim$ 100 -- 700 \kms) and more than two
orders of magnitude in X-ray luminosity (Beers \etal\ 1995, Kriss \etal\ 1980,
1983).  Detailed study of these systems may thus discriminate among
these scenarios.  These systems are dynamically fairly simple
in their centers, without much substructure in the central $1 h^{-1}$ Mpc,
although there is evidence for complexity on larger scales (Beers \etal\ 1995).
They are sufficiently nearby that a complete
sample can be analyzed to reasonable limiting magnitude with
modest instrumentation.  X-ray data in conjunction with optical studies
provide a foundation for equilibrium mass models of these clusters.

Early optical studies of AWM/MKW systems lack extensive photometry and complete
redshift samples (Beers \etal\ 1984, 1995 and references therein;
Malumuth \& Kriss 1986; Williams \& Lynch 1991, Price \etal\ 1991;
Dell'Antonio \etal\ 1995).  AWM~7 is particularly problematic because it
spans two Palomar Sky Survey plates with rather different sensitivities,
making consistent photographic magnitude determination difficult.
Previous optical determinations of the mass-to-light ratio of AWM~7 
(Kriss \etal\ 1983, Beers \etal\ 1984) yield
values exceeding 1000 $h$ $M_\odot/L_\odot$, a result exacerbated
by the cluster's high velocity dispersion.  The optical data are discrepant
with X-ray 
determinations in the range 200--500 $h$ $M_\odot/L_\odot$ (Dell'Antonio
\etal\ 1995, Neumann \& B\"ohringer 1995).  Our new complete photometric
and spectroscopic data yield $\sim$$650 \pm 150$$h$ $M_\odot/L_\odot$,
consistent with the range for other clusters and with the X-ray data for
AWM~7.  Optical mass-to-light determinations for other AWM/MKW systems
by Beers \etal\ (1995) are generally consistent with X-ray results.

AWM 7 is the first system in our
study of a complete sample of nearby AWM/MKW clusters.  It is one of the
nearest clusters ($\sim$5200 \kms) and is the one with the largest velocity
dispersion ($\sim$700 \kms).
We have measured 172 magnitudes and redshifts in AWM~7, making it one
of the best-sampled systems in the sky.

In \S 2, we discuss the data acquisition and reduction, and define the
cluster sample.  In \S 3, we discuss the cluster kinematics and segregation
by spectral type, search for substructure, and examine the velocity
dispersion profile.  In \S 4, we discuss the photometric properties,
compute the luminosity function of the cluster, and compute the mass-to-light
ratio.  We discuss the ramifications in \S 5 and conclude in \S 6.

\section{Observations}

AWM 7 is a poor cluster with the cD galaxy NGC~1129 ($\alpha =$2:54:27,
$\delta = $41:34:47 J2000) at its center.  The recession velocity
of the cD is $5288\pm 71$ \kms. The cluster has a mean radial
velocity $5247\pm 76$ \kms.  Here we describe our observations of the
system, and define the cluster sample.

We measured redshifts of 172 galaxies
within a projected distance of 1.6 $h^{-1}$ Mpc from NGC 1129, and 
obtained $R$-band CCD photometry for the central $1.15 \times 1.15$
$h^{-1}$ Mpc of the cluster.  127 of the galaxies with redshifts are in
the region with CCD photometry.

Table~\ref{tab:datasample} contains the
velocities and $R$-band magnitudes for the 172 galaxies.
Column 1 lists the galaxy RA; column 2, the declination; column 3, the radial
velocity; column 4, the uncertainty in the radial velocity; column 5, the
isophotal magnitude (to 23.5 mag asec$^{-2}$ in $R_{\rm KC}$); column 6, the
error in isophotal magnitude; column 7, the source of the magnitude (CCD or
POSS); column 8, the spectral type (presence or absence of H$\alpha$ emission,
quantified in \S 3); and column 9, the extinction $A_R$
along the line of sight to the galaxy.

\subsection{Redshifts}

We measured 140 redshifts with the FAST spectrograph on the Whipple Observatory
1.5-m Tillinghast telescope in Nov.--Dec. 1995 and a further 43 with the MMT
Blue Channel Spectrograph on 1996 Dec. 5--6; yielding 172 galaxy redshifts.
To ensure uniform spectroscopy with a well-understood error model, we obtained
a new spectrum for each galaxy despite some previous measurements in the
literature ($e.g.$ Beers {\it et al.} 1995 and references therein).  Kurtz \&
Mink (1998) discuss in detail how FAST velocities compare to other measurements;
briefly, the instrumental velocity offset with respect to the night sky is
less than 10 km~s$^{-1}$ and is indistinguishable from zero.
The FAST spectra have 6 \AA\ resolution and spectral coverage of 3600--7600 \AA.
The MMT spectra have 10.6 \AA\ resolution and coverage of 3000--8000 \AA.
We reduced the spectra by cross-correlation, using the IRAF XCSAO task
(Kurtz \& Mink 1998).

Most galaxies in our survey are absorption-line systems.  For these,
the mean uncertainty in the measured velocities is $\sim$35 \kms.
For the emission-line galaxies
we add an extra error term of 60 \kms in quadrature because the line-emitting
regions do not necessarily trace the galaxy center of mass (Kurtz \& Mink
1998); the values tabulated in Table~\ref{tab:datasample} incorporate this
extra term.

We used the Digitized Palomar Observatory Sky Survey scans to select
targets for observation with FAST within a $2^\circ \times 2^\circ$ region
centered on NGC 1129.  AWM~7 spans two POSS plates that
differ greatly in sensitivity.  Uniform magnitude determinations are thus
difficult; the problem is compounded by vignetting at the plate edges.
On the basis of the CCD photometry we later defined a sample complete to
$R = 16.5$ covering the central 
$75' \times 75'$ ($1.15 \times 1.15$ $h^{-1}$ Mpc).  We
used the MMT to measure redshifts of galaxies with $R \gtrsim 16.0$.

\subsection{Photometry}

We acquired an $R$-band mosaic of the central $75'\times 75'$ of the cluster
with the MDM 1.3-m telescope during Nov. 1995.  There are 36 six-minute
$13.7' \times 13.7'$ exposures, with $1.2'$ overlap between frames.
Conditions were photometric
for most of the images; we correct the 4 non-photometric images using
the overlap regions with neighboring photometric frames.  The rms error
in the photometric solutions are 0.0196 and 0.0370 mag for the two photometric
nights.  We determine isophotal magnitudes with the FOCAS package.  The
quoted magnitudes are isophotal to $R = 23.5$ mag/arcsec$^2$, which is 
more than $2\sigma$ above the sky noise in each frame.  We reviewed the
star/galaxy separation manually for all non-stellar objects brighter than
$R_{23.5} = 18.5$ to remove misclassified objects --- FOCAS tends to misclassify
double stars as galaxies.

FOCAS magnitudes are sensitive to the local sky.  FOCAS is optimized for
faint objects, and tends to underestimate the luminosity of bright extended
objects systematically, assigning much of the diffuse light to the sky rather
than to the object.  Moreover, because the isophotes are defined by the 
number of $\sigma$ in excess of sky, the determination of the sky
$\sigma$ in each frame is critical.  There are flaws in the FOCAS software
which lead to systematic errors in the sky determination.  The presence of 
bright stars at the bottom edge of an image results in comet-like
swaths of spuriously bright sky extending up the image.  Blocking out the
stars by hand removes this effect, but its very existence adds uncertainty
to the sky determination, and hence to the isophotal magnitudes.
We thus add 0.05 mag in quadrature to the magnitude errors, derived
otherwise from the rms scatter in the photometric solution; for the
non-photometric images, we also include the 
uncertainty in the zeropoint as determined
from the rms magnitude error for isolated stars in the overlap regions.
We quote an error of 0.25 mag for the Digitized POSS magnitudes (Geller
{\it et al.} 1997), although the scatter is in fact likely greater
for AWM~7 because of the different sensitivities of the two plates spanned
by the cluster.

\begin{figure}[p]
\centerline{\epsfxsize=6.0in%
\epsffile{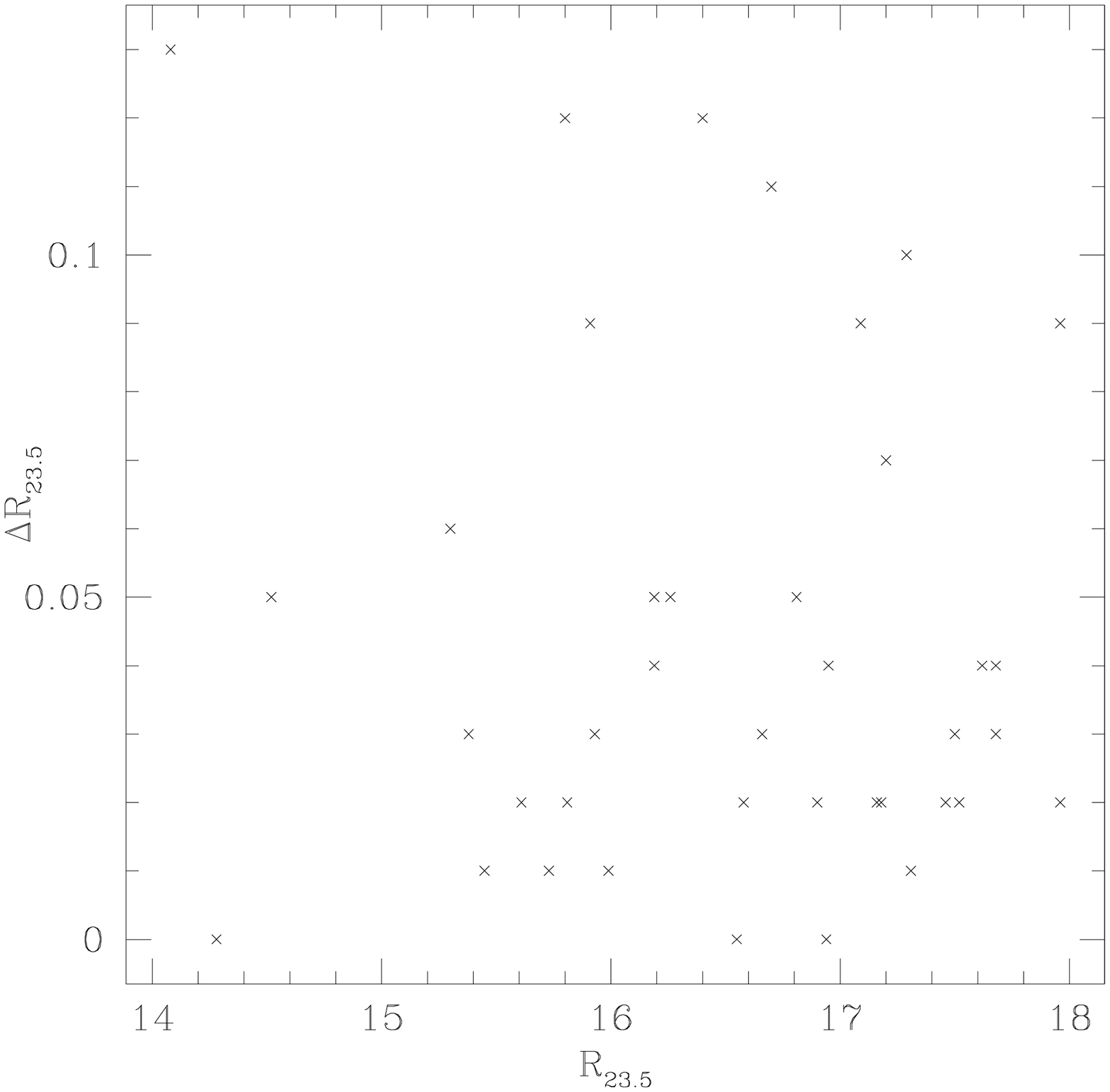}}
\caption{Absolute value of the difference in magnitudes determined from
separate mosaic fields, for galaxies (in overlap regions) with multiple
magnitude measurements.  Each such galaxy generates one point on this
plot; we plot the greatest magnitude difference against the brightest
magnitude.}
\label{fig:magconsist}
\end{figure}

We use galaxies in overlap regions of the mosaic to check the
consistency of the magnitudes; in Fig.~\ref{fig:magconsist} we plot the
difference in the magnitude measurements for individual galaxies as
determined from different mosaic fields.  We plot the differences as a
function of the brighter magnitude.  When a galaxy appears in more than
two fields, we plot the greatest difference against the brightest
magnitude.  The mean difference is 0.043 mag, with a scatter of 0.036 mag.

We calculate an extinction correction for each galaxy from the
relation $A_R = 2.5 E(B-V)$ (Zombeck 1992), with the color related
to the H{\sc i} column density by $\left< N({\rm H {\sc i}})/E(B-V) \right> =
4.8 \times 10^{21}$ atoms cm$^{-2}$ mag$^{-1}$ (Bohlin \etal\ 1978).
We obtain the H{\sc i} column density along the line of sight to each
galaxy from the Bell Labs H{\sc I} maps (Stark \etal\ 1992).
The extinction in the total sample of
172 galaxies ranges from 0.40 to 0.54 mag; the gradient arises from the
cluster's proximity to the galactic plane ($b_{\rm cD} = -15\fdg6$).  Within
the region with CCD photometry, the extinction ranges from 0.42 to 0.50.
These values exceed those on the map of
Burstein \& Heiles (1982) by $\sim$0.15 mag.  The magnitudes
listed in Table~\ref{tab:datasample} are the measured magnitudes,
uncorrected for the extinction in column~(9).  Calculations of
the luminosity function below do account for the extinction.

\subsection{Defining the cluster sample}

\begin{figure}[p]
\centerline{\epsfxsize=6.0in%
\epsffile{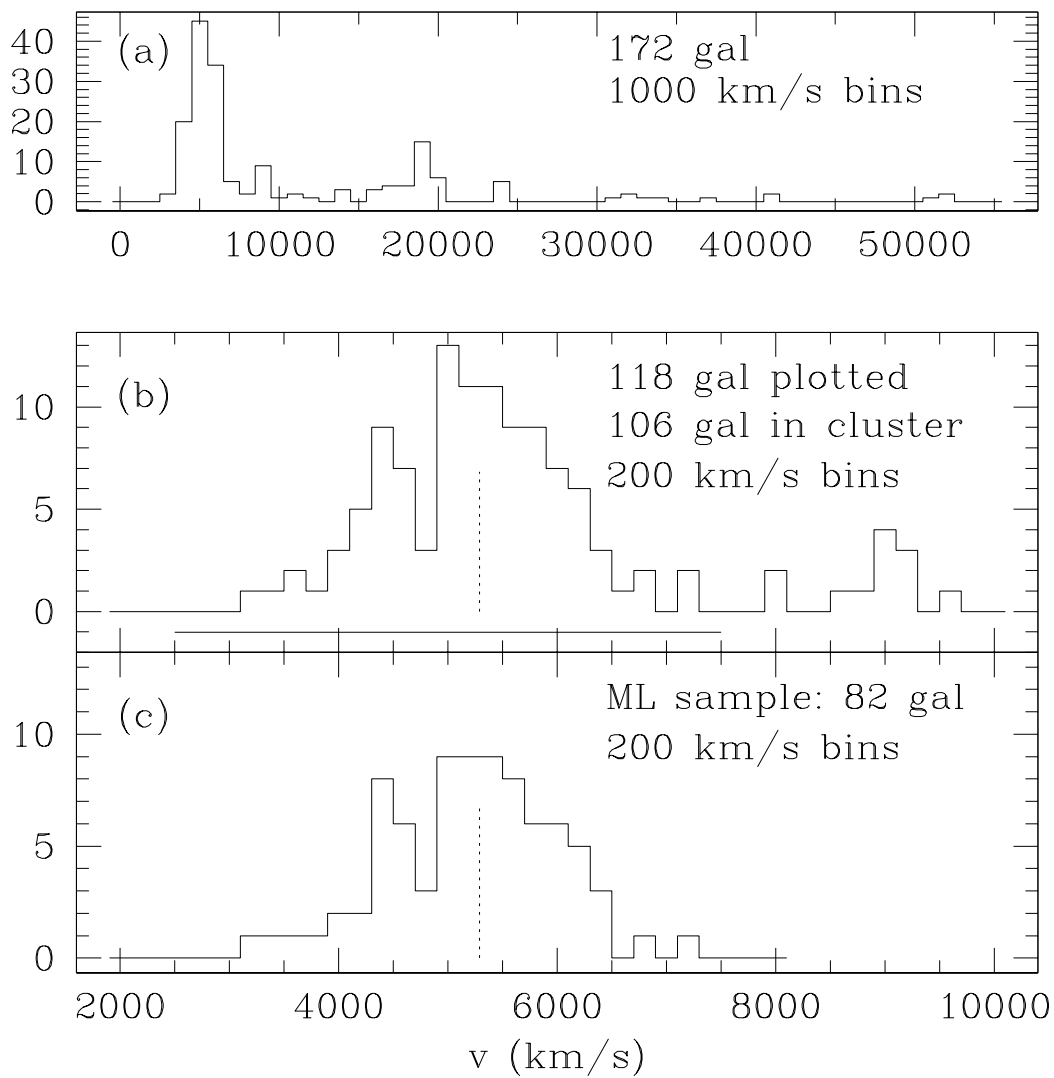}}
\caption{(a) Velocity distribution of the 176 galaxies with measured redshifts.
(b) Velocity distribution of galaxies with $cz<10000$ \kms.  The solid rule
beneath the histogram indicates the velocity criterion for cluster membership.
(c) The velocity distribution of the 82 galaxies with $2500<v<7500$ \kms
and with CCD photometry that constitute the core sample.
In (b) and (c) the velocity of the central cD galaxy is indicated by the
dotted line.}
\label{fig:vhistmulti}
\end{figure}

Figure~\ref{fig:vhistmulti}a shows the redshift distribution 
for all 172 galaxies; large-scale structure is apparent behind the cluster
at $\sim$20000 \kms, and more weakly at $\sim$10000 \kms.  In the range
2500--7500 \kms there are 106~galaxies, which we identify as cluster members;
we denote this set of galaxies the ``C'' sample.
The line under the histogram in Figure~\ref{fig:vhistmulti}b
indicates this range and the dotted line indicates the redshift of the
central cD galaxy (5288$\pm$71 \kms).  For these 106 galaxies, 
$c\bar z = 5247 \pm 76$ \kms and  $\sigma = 783^{+60}_{-49}$ \kms
(Danese {\it et al.} 1980, 68\% confidence).
There is a 779 \kms velocity gap between the highest-redshift
cluster galaxy and the lowest-redshift background galaxy; this is a 
$\sim$1$\sigma$ gap starting 2.5$\sigma$ above the cluster mean.  There is
no obvious foreground.

82 of the 106 C galaxies lie within
the region with CCD photometry; Figure~\ref{fig:vhistmulti}c shows their
velocity distribution.  This sample is 100\% complete to $R=16.3$, 99\% complete
to $R=16.5$, 98\% complete to $R=16.7$, and 96\% complete to $R=16.9$
(uncorrected magnitudes); for this subsample,
$c\bar z = 5248 \pm 82$ \kms and $\sigma = 747^{+66}_{-52}$ \kms, clearly
consistent with the larger sample.  We denote this set of galaxies the
``ML'' sample.

There are 134 galaxies in the central $75' \times 75'$ (1.56 deg$^2$) 
with $R_{23.5} \leq 17$, of which 122 have measured redshifts; of these, the 82
ML galaxies have velocities in the range 2500--7500 \kms.  Assuming that
82/122 of the
12 unmeasured galaxies also lie in the cluster, we estimate that there are
$4 \pm 2$ additional background galaxies, for a total of $44 \pm 5$, or
$28 \pm 3$ per square degree.  Representing the background count by
$$n_b = C_0 \int_{-\infty}^{m_{\rm lim}} 10^{d_0m} \, dm \quad {\rm deg}^{-2}$$
and using values of $d_0$ and $C_0$ derived from the Century Survey
(Geller {\it et al.} 1997) yields
$n_b = 36 \pm 6\ {\rm deg}^{-2}$ for $m_{\lim} = 16.9$, consistent with the
background we observe.  The quoted errors are Poisson errors, which
underestimate the true error due to clustering.

\begin{figure}[p]
\centerline{\epsfxsize=6.0in%
\epsffile{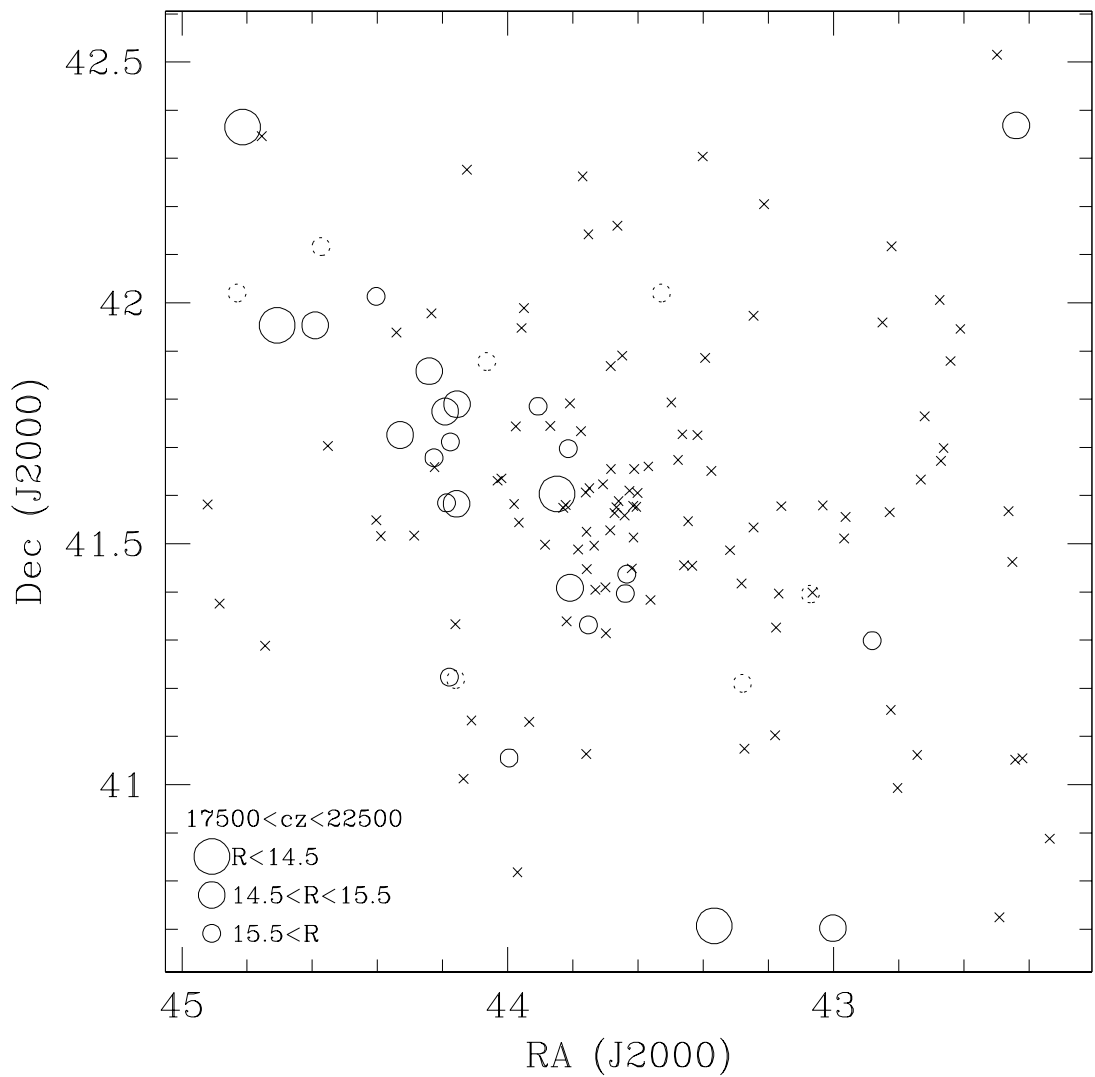}}
\caption{The angular distribution of the clusters and background structure.
Crosses are galaxies
in AWM~7, with $2500<cz<7500$ \kms.  Dashed circles are background galaxies
with $15000<cz\leq 17500$ \kms.  Solid circles are background galaxies with
$17500<cz<22500$ \kms, coded by size for magnitude.}
\label{fig:twoclustspmag}
\end{figure}

We investigate the peak in the velocity histogram (Fig.~\ref{fig:vhistmulti}a)
at $\sim$18,000 \kms.
We plot the spatial distribution of the background galaxies
in Fig.~\ref{fig:twoclustspmag}; the non-uniform distribution of the
background galaxies adds some uncertainty to the computation of the faint end
of the LF (\S 4.3) and to the background-subtraction statistics.

\section{Kinematics}

We use the C sample to examine the kinematics of the cluster.  This sample
is magnitude-limited only within the area of the CCD survey.
We separate our sample by spectral type (presence/absence of H$\alpha$);
the two subsamples have quite different kinematics.  We test for substructure
in the cluster and examine the cluster velocity dispersion profile as a 
function of radius.

\subsection{Velocity Histogram}

The velocity distribution in Figure~\ref{fig:vhistmulti}b appears 
bimodal with an apparent peak near 4500 \kms.  However, a Kolmogorov-Smirnov
test (Press \etal\ 1992) shows that the distribution is consistent with
a Gaussian velocity distribution of mean 5247 \kms and dispersion 783 \kms
($P_{D > D_{\rm obs}} = 0.73$).
 
\subsection{Spectral Segregation}

We separate the sample into emission (Em) and non-emission (Ab) galaxies,
based on the presence or absence of H$\alpha$ emission in the spectrum.
We use two criteria for including a galaxy in the Em sample.  The first is
that the redshift derived from cross-correlating against the emission-line
template lie within 200 \kms of the redshift derived from the best-fit
template; if the emission-line template fits best, this criterion is
automatically satisfied.  The second criterion is that the EMSAO task in
the IRAF RVSAO package must detect and correctly identify
H$\alpha$ given the best-fit redshift.  If both criteria are satisfied,
we classify the galaxy as Em; if neither, as Ab.  If the first criterion is
satisfied and not the second, the galaxy is classified as Ab; given the
correct redshift, EMSAO would identify any strong H$\alpha$ that were present.
If the second criterion is satisfied but not the first,
we inspect the spectra visually; most such galaxies are ultimately classified
as Em.  The robustness of the classification
is clearly a function of the signal-to-noise ratio of the spectrum; emission
line galaxies with poor signal-to-noise or only weak emission may be
misclassified as non-emission if the H$\alpha$ emission line
does not rise appreciably above the noise.
For Em galaxies with strong emission lines and high S/N spectra, the
emission-line H$\alpha$ equivalent widths are typically $\gtrsim$ 7 \AA;
these are indicated with an asterisk in Table~\ref{tab:datasample}.
The equivalent widths in the other Em galaxies (those with intrinsically
weaker lines, or just with lower S/N spectra) range from 2 to 7 \AA\ .
Of the 172 galaxies observed, 50 show emission and 122 do not; of the 82
galaxies in the ML sample, only 9 show H$\alpha$ emission.
 
\begin{figure}[p]
\centerline{\epsfxsize=6.0in%
\epsffile{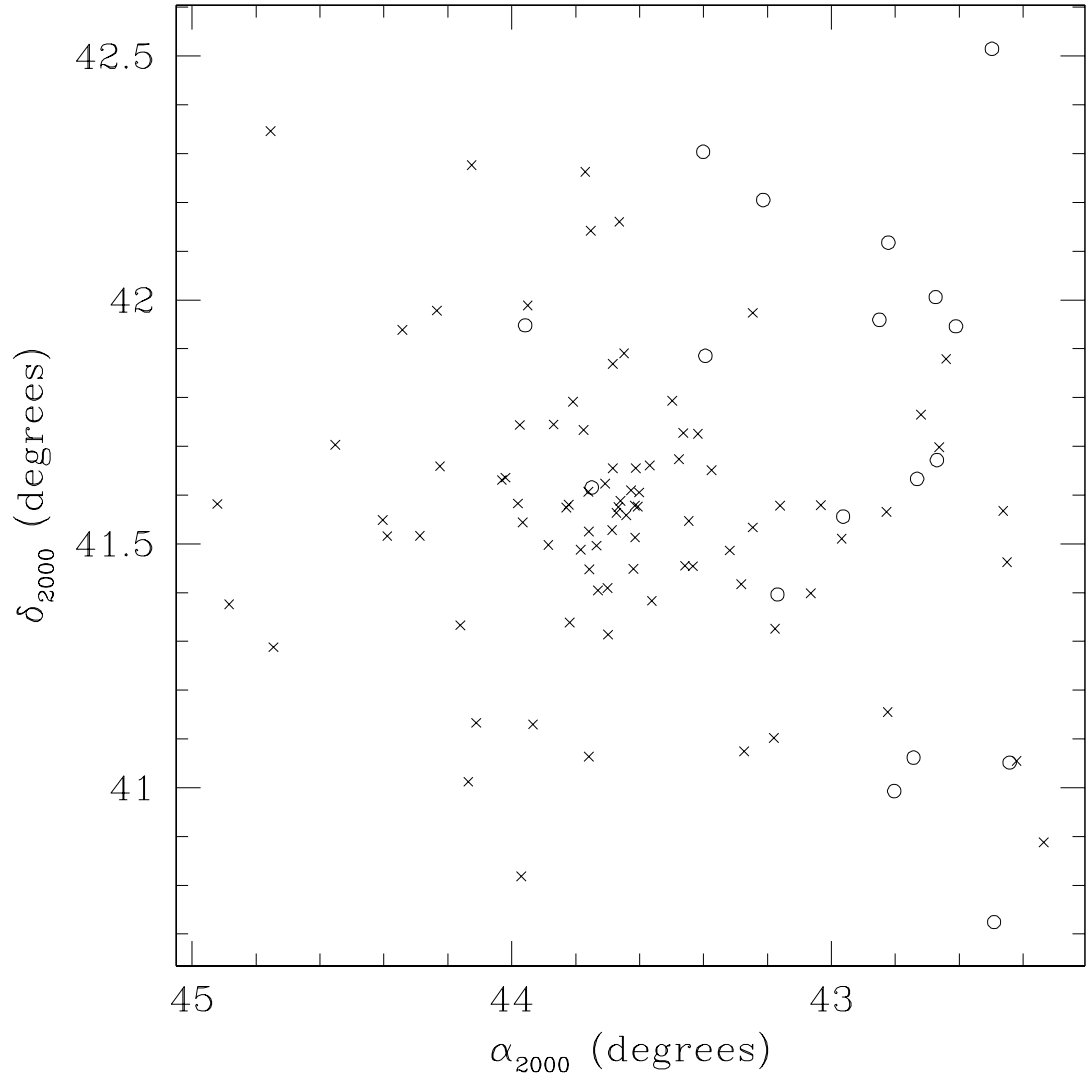}}
\caption{The angular distribution of the 106 C galaxies with $2500<v<7500$ \kms,
both within and without the region with CCD photometry.  Emission-line galaxies
are circles; non-emission galaxies are crosses.}
\label{fig:spatspec}
\end{figure}

The Em and Ab galaxies are spatially segregated.
Fig.~\ref{fig:spatspec} shows the distribution of 106 galaxies for which
2500 \kms $< cz <$ 7500\kms, with emission-line galaxies
plotted as circles and non-emission galaxies as crosses.  The
distribution of emission-line galaxies is not centrally concentrated,
nor is it centered on the cD.  The median distance of the Em galaxies
from the central cD galaxy is 44\farcm6; for the Ab galaxies it is 17\farcm8.

\begin{figure}[p]
\centerline{\epsfxsize=6.0in%
\epsffile{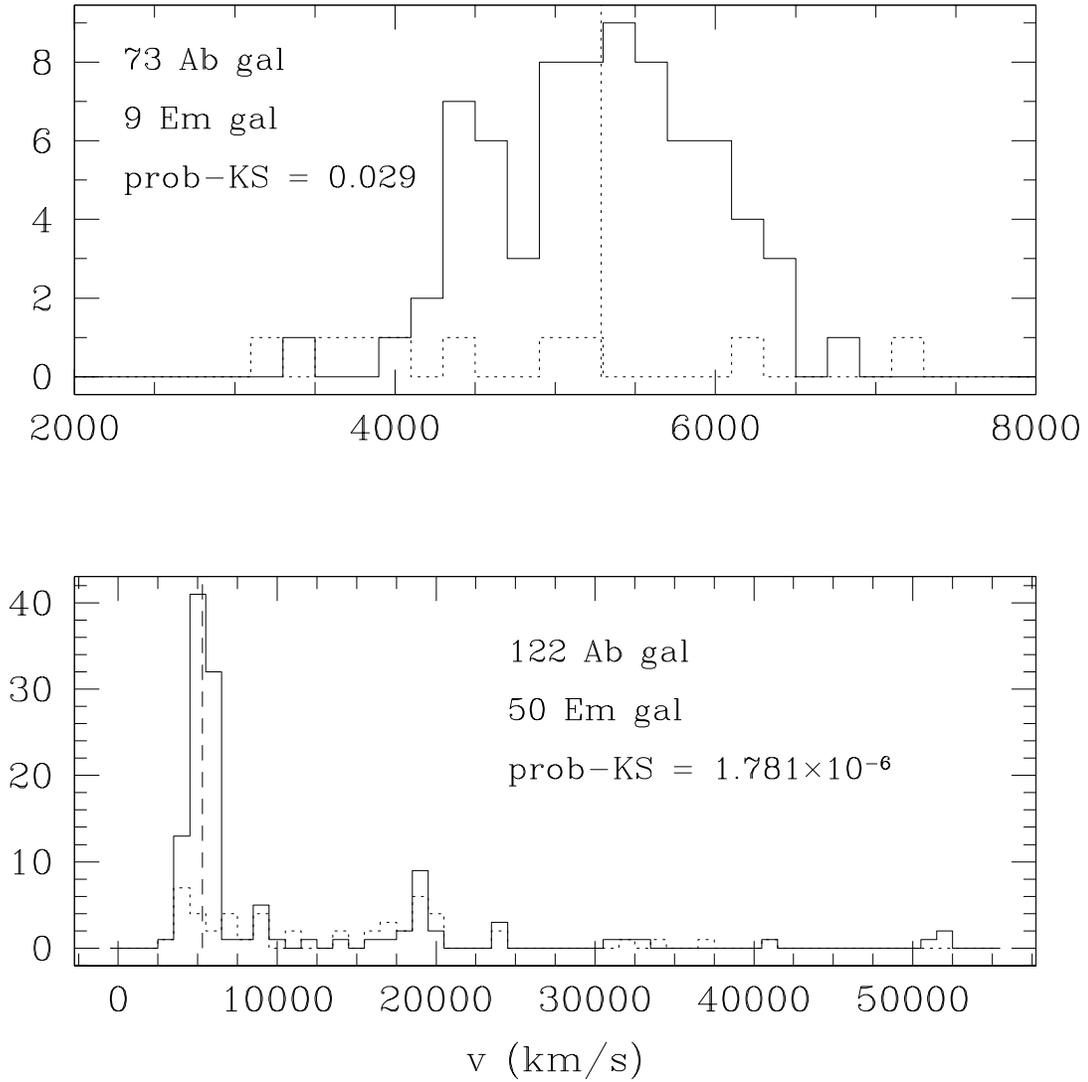}}
\caption{Velocity distribution of the emission and non-emission galaxies.
The upper panels show only galaxies in the ML sample with CCD photometry.
The central cD's velocity is indicated with a dotted line in each panel.}
\label{fig:vhistspec}
\end{figure}

The velocity and magnitude distributions of the emission-line and
non-emission galaxies also differ.  Fig.~\ref{fig:vhistspec} shows the
velocity distribution of the Em and Ab galaxies separately for the ML 
sample in the upper panel, and for all observed galaxies in the lower
panel.  The 9 emission-line galaxies have
$c\bar z = 4971 \pm 438$ \kms, $\sigma = 1313^{+503}_{-233}$ \kms;
the 73 non-emission galaxies have
$c\bar z = 5304 \pm 75$ \kms, $\sigma = 643^{+61}_{-48}$ \kms.  The
velocity distribution of the Em galaxies is apparently much broader,
but because of the small size of the sample, a two-sample K-S test rules
out the Em and Ab galaxies' being drawn from the same underlying velocity
distribution with only 97\% confidence.

\begin{figure}[p]
\centerline{\epsfxsize=6.0in%
\epsffile{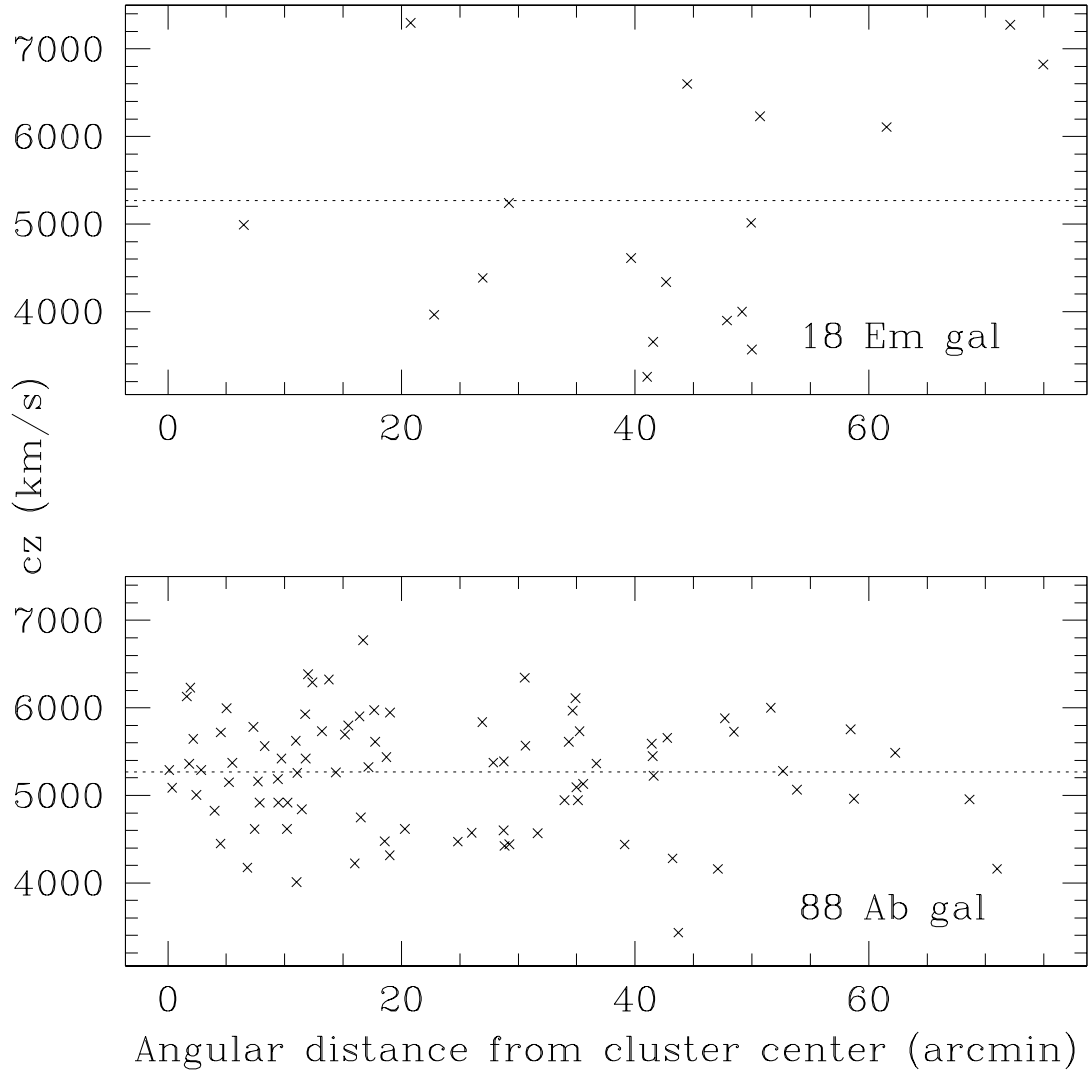}}
\caption{Radial velocity as a function of angular distance from the central cD
for the Em and Ab galaxies in the C sample.  The dotted line indicates the mean
cluster redshift.}
\label{fig:vofr}
\end{figure}

Figure~\ref{fig:vofr} shows velocity as a function of angular distance from
the cluster for all 106 C galaxies.  The greater central concentration of
the Ab galaxies is apparent, as is their smaller velocity dispersion.
Moreover, all but one of the 13
faint ($R>16.3$) cluster galaxies are absorption line systems.
We discuss the magnitude distribution in more detail in \S 4.

The cluster Abell~576 shows similar behavior (Mohr {\it et al.} 1996a); there
too the Em galaxies are less spatially concentrated, have a greater
velocity dispersion, and are systematically fainter than the Ab galaxies,
but they are not offset from the cluster center.  The core velocity dispersions
of the Ab galaxies are $\sim$530 \kms in both AWM~7 and A~576, but the
velocity dispersion profile rises more steeply and to a higher value in A~576;
at 1 Mpc, $\sigma_{\rm A 576} \sim 1000$ \kms.
The ratio of Em to Ab galaxies is larger in A~576 (79:142).  We follow the
procedure adopted by Mohr {\it et al.} (1996a) and consider the Em and Ab
samples separately below.  We base our estimates of the mass-to-light ratio on 
the Ab galaxies only.

\subsection{Substructure}

We use the Dressler-Shectman statistic (Dressler \& Shectman 1988) to test
for substructure in the cluster.  They define the statistic $\Delta_0 =
\Sigma_i \delta_i$, where the summation is over all galaxies and
$$
\delta_i \equiv {{n}\over{\sigma_{\rm g}^{2}}}
           [ (\bar v_{\rm g} - \bar v_i)^2 +
             (\sigma_{\rm g} - \sigma_i)^2 ]^{1/2}
$$
is a measure of the deviation of the local mean velocity and
dispersion ($\bar v_i$, $\sigma_i$) from the global cluster values
($\bar v_{\rm g}$, $\sigma_{\rm g}$).  For each
galaxy, $\delta_i$ is a function of the number of nearest neighbors $n$
entering into the calculation of the local $\bar v_i$ and $\sigma_i$.
We evaluate the significance of $\Delta_0$ for each $n$ by randomly shuffling
the velocities of all galaxies 5000 times, and re-calculating $\Delta_0$
each time.  We thus obtain a distribution of $\Delta_0$ against which to
compare the actual value.

\begin{figure}[p]
\centerline{\epsfxsize=6.0in%
\epsffile{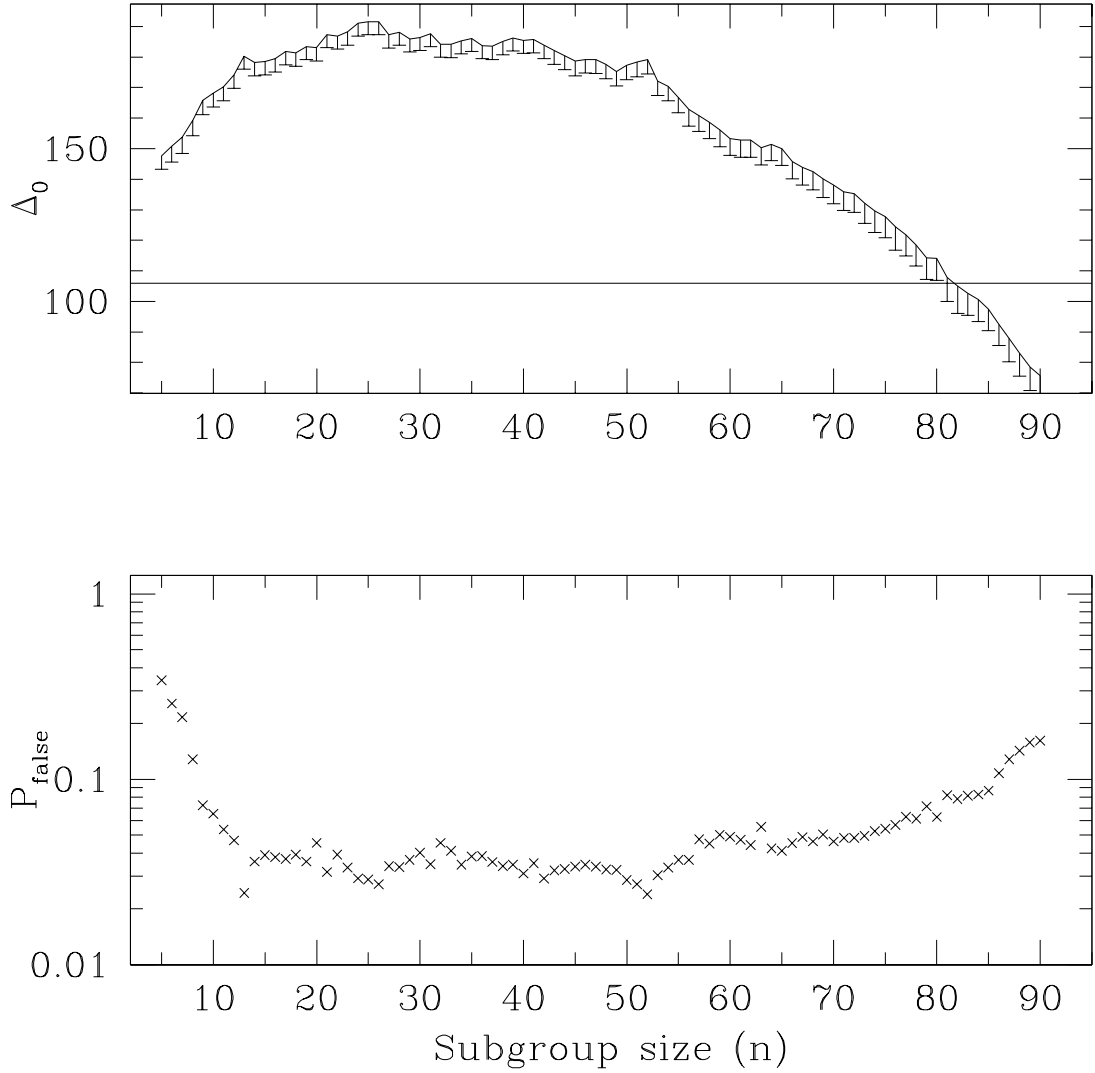}}
\caption{Top: the Dressler-Shectman statistic $\Delta_0$ as a function of
subgroup
size $n$.  The horizontal rule is at $\Delta_0 = 106$, the (Em+Ab) sample size.
Bottom: the probability of $\Delta_0$ being equal to or greater than
the observed value by chance, as derived from 5000 Monte Carlo simulations.}
\label{fig:dressdelta.en}
\end{figure}

A D-S test with $n=11$ 
indicates that there is substructure in the northwest of the cluster, where the
Em galaxies predominate.  Because the $\Delta_0$ statistic characterizes
local deviations of the mean and dispersion from
the overall cluster values, this substructure reflects the larger dispersion
of the Em galaxies seen in Fig.~\ref{fig:vhistspec}.  The
significance of the substructure detection is marginal.
Figure~\ref{fig:dressdelta.en} shows the
D-S statistic for the 106 C galaxies as a function of subgroup
size $n$ for $5\leq n\leq 90$ in the
top panel, with the probability of an equal or greater D-S statistic
arising by chance (determined from the 5000 Monte Carlo simulations for
each $n$) in the lower panel.  A low $P_{\rm false}$
indicates a high significance for the substructure detection.
The substructure is most significant for
$n=13$, but even then there is still a greater than
2\% chance of an equal or greater $\Delta_0$ arising by chance.
When we exclude the Em galaxies and perform the D-S test
on the remaining 88 Ab galaxies, there is no discernible substructure
for any  value of $n$ (not shown); 
the distribution of Ab galaxies is smooth.  This analysis supports the
idea that the Em galaxies are a dynamically distinct population of
late-type galaxies; our sample does not contain enough Em galaxies  to
determine their large-scale dynamics.  It may be that, as in A~576, they
are infalling.

\subsection{Velocity Dispersion Profile}

\begin{figure}[p]
\centerline{\epsfxsize=6.0in%
\epsffile{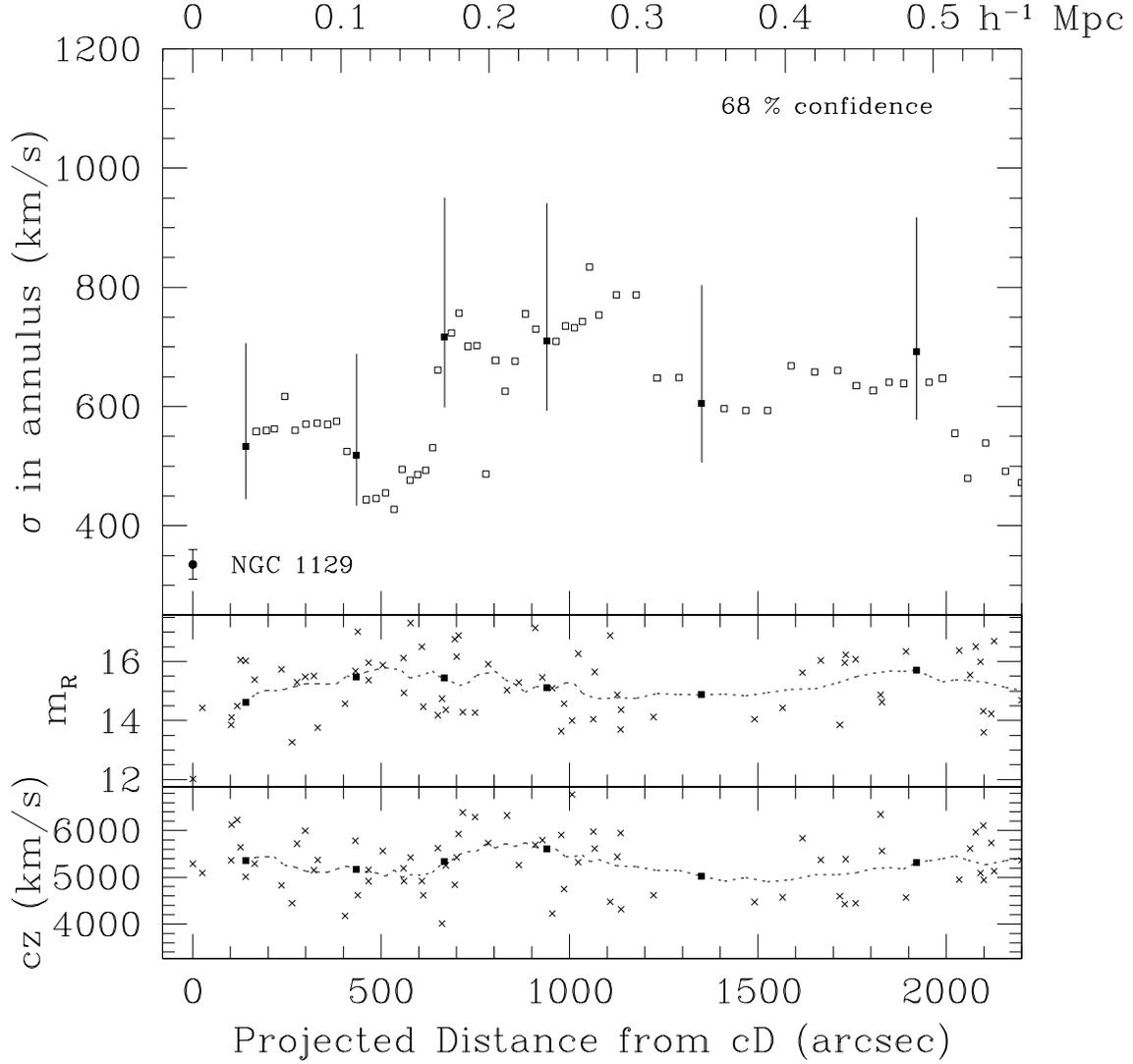}}
\caption{Velocity dispersion, apparent magnitude, and velocity of Ab-type
galaxies as a function of projected cluster
radius.  $\sigma$ and median values of $m_R$ and $cz$ (dotted lines) are
calculated from a sliding bin of 11 galaxies; independent
(uncorrelated) points are filled squares, with 68\% error bars for $\sigma(r)$.
Data point for N1129 represents internal $\sigma$ of galaxy.  In the lower
panels, crosses denote individual galaxies.}
\label{fig:sigmavsr2}
\end{figure}

Figure~\ref{fig:sigmavsr2} plots the velocity dispersion of the 88
non-emission C galaxies as a function of cluster radius.  We take the cD
as the geometric center of the cluster; Neumann \& B\"ohringer (1995,
hereafter NB) find that the cD coincides exactly with the maximum of
the X-ray emission.  The plot extends to to 2200 arcsec, the radius to which
our photometric and kinematic data are both complete.

Each point in the upper panel of Fig.~\ref{fig:sigmavsr2} represents
the velocity dispersion of 11 galaxies ranked sequentially in distance from
the cD; neighboring points are thus correlated, but represent annuli of
different widths.  Uncorrelated points are distinguished by 68\%
confidence-level error bars. The lower panels plot the
velocity and magnitude of the galaxies along with a moving median curve;
uncorrelated points are indicated by filled boxes along the curve.
The data point for NGC~1129 is a measurement of its internal velocity
dispersion from Malumuth \& Kirshner (1985), with a $4'' \times 10''$ aperture.
The internal $\sigma = 335\pm 25$ \kms is substantially less than the
velocity dispersion in the cluster core.

The median velocity throughout the core strays little from the overall cluster
median; However, the cluster velocity dispersion appears lower in the core.
Within $0.1 h^{-1}$ Mpc of the center, $\sigma \sim 550_{-100}^{+150}$ \kms,
some 200 \kms lower than at the periphery; about 20 galaxies comprise
the cold core.  To assess the significance of the cold core, we
split the sample by radius and evaluate $\sigma$ separately for galaxies within
and external to the delimiting radius.  This analysis indicates that the core is
cooler than the outskirts, but only at the $\sim$$1\sigma$ level.
The evidence for a cold core from $\sigma(r)$ alone is thus present, but weak.
The scale of the cold core matches the scale of the X-ray cooling flow seen
by NB.

Diaferio (1997) proposes a simple dynamical explanation for a cold core.
Under the assumption of virial equilibrium
(probably valid for the Ab-type galaxies in the core),
$$\sigma^2 \propto { {GM(<r)} \over {r} } \quad .$$
Representing the radial mass density profile by a power law
$\rho(r) \propto r^{-\alpha}$ yields
$$ M(<r) \propto \int_0^r \rho(x) x^2 \, dx 
   \propto r^{3-\alpha} \quad ,$$
so $\sigma \propto r^{1-\alpha/2}$.  Thus $\alpha < 2$ results in a rising
$\sigma(r)$ profile, and $\alpha > 2$ in a falling profile.  The mass model
of Navarro \etal\ (1995) posits $\rho(r) \propto r^{-1}(r+r_s)^{-2}$,
which behaves as $r^{-1}$ for small $r$, giving $\sigma(r) \propto r^{1/2}$,
a rising profile.  Alternatively, the common $\beta$-model
(Cavaliere \& Fusco-Femiano 1978) has
$\rho(r) \propto [1 + (r/r_c)^2 ]^{-3\beta / 2}$, implying constant density
for $r \ll r_c$, and so $\sigma(r) \propto r$ in this regime.
NB find $r_c = 51 \pm 3 h^{-1}$ kpc for AWM~7.  Thus the $r$-dependence of 
$\sigma$ in the core ($r < r_c$) would only be detectable 
with very dense sampling of the cluster core to overcome
small-number Poisson statistics, the sensitivity of $\sigma$ to outliers, and
the small angular extent of the core.
Our sampling is too sparse to characterize any rise within $r_c$ as more
representative of one model density profile or the other;
a deeper sample could in principle discriminate between them.  However, the
cluster core has relatively few faint galaxies, and the core sampling may
never be dense enough to discriminate.

\section{Photometric Properties}

\begin{figure}[p]
\centerline{\epsfxsize=6.0in%
\epsffile{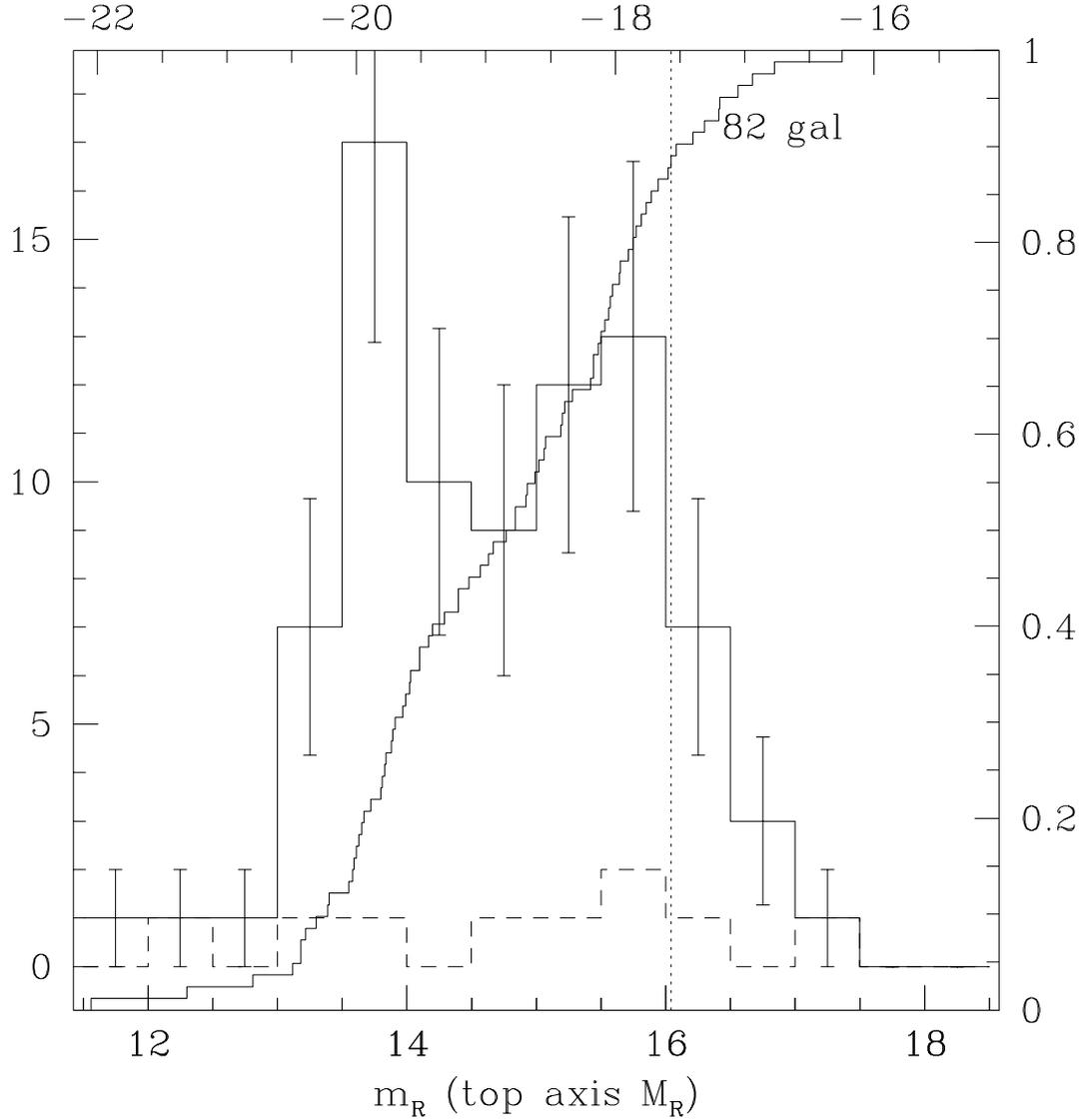}}
\caption{The differential (left-hand scale) and cumulative (right-hand scale)
extinction-corrected apparent magnitdue distributions for the ML sample.
Errors on the histogram are $\sqrt{N}$.}
\label{fig:sampmaghist}
\end{figure}

Figure~\ref{fig:sampmaghist} shows the differential and cumulative
magnitude distribution of the 82~ML galaxies, with the Em galaxies alone
as the dotted histogram.  Here we correct the magnitudes
for extinction.  At the mean sample redshift of 5247 \kms,
$m=M + 33.60 - 5\log h$.  A fiducial absolute magnitude for the field 
$M_{*F} = -20.7$ in the $R$-band from the Century Survey
(Geller {\it et al.} 1997) yields
a corresponding $m_{*F} = 13.35 - 5\log h$.   AWM~7 contains three galaxies
substantially brighter than $m_{*F}$, and seven of comparable magnitude.  

The distribution of the Em galaxies is flat as a function of magnitude;
significantly, only one galaxy fainter than 16.3 is an emission-line galaxy.
The median magnitude of the Em sample is $R=15.44$;
the median of the Ab sample is $R=14.67$.
The offset between the magnitude distributions probably reflects the
$\Delta(B-R) \sim 1$ mag color difference between late- and early-type
galaxies.  At $B$, the Em and Ab galaxies would have more concordant magnitude
distributions, consistent with field measurements (Marzke \etal\ 1994).
A similar offset between the spectral types is seen in A~576.

\subsection{Surface Brightness}

\begin{figure}[p]
\centerline{\epsfxsize=6.0in%
\epsffile{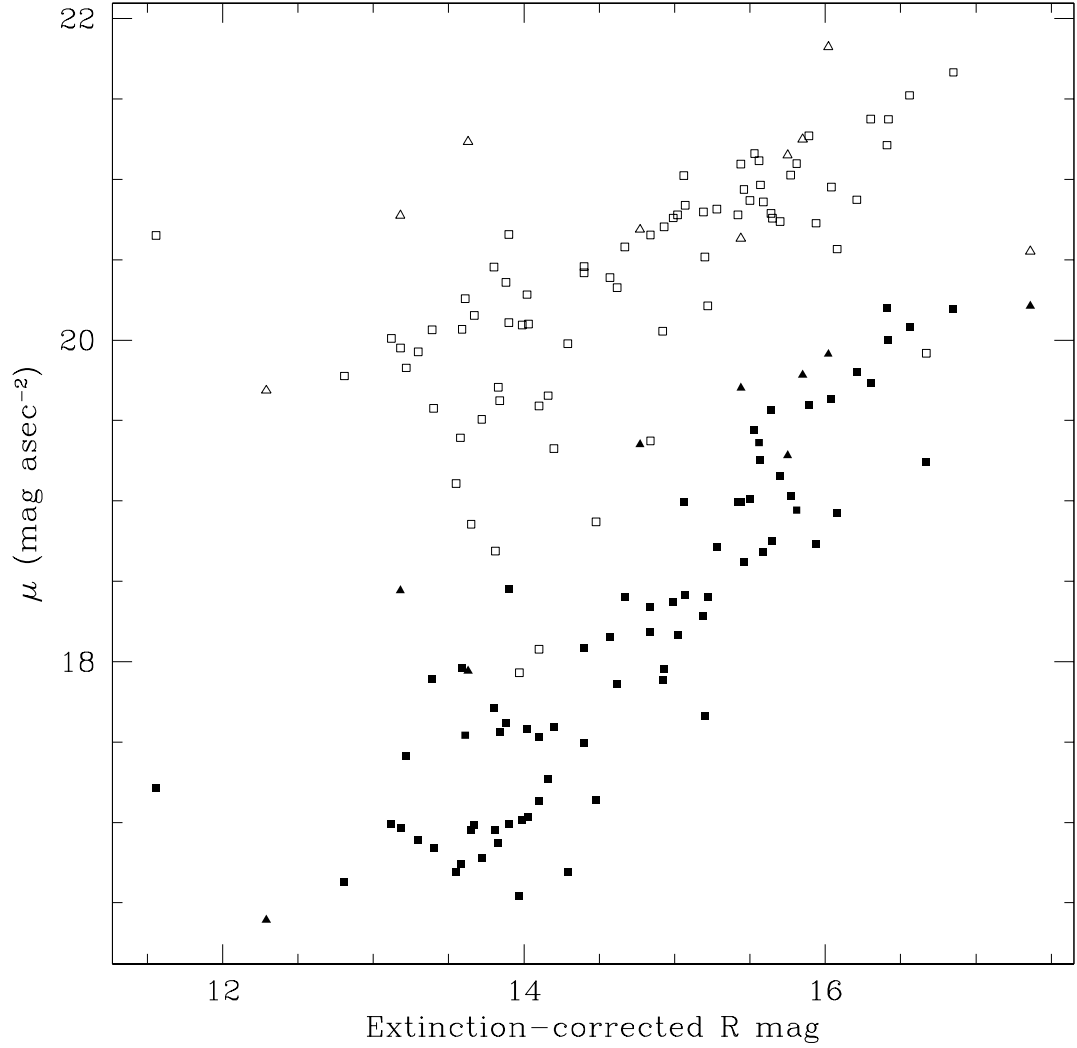}}
\caption{Mean (open) and central (solid) surface brightness for the 82
sample galaxies.  The central surface brightness is determined in the
3$\times$3 pixel grid with the highest flux, corresponding to a
patch 1$\farcs$33 square on the sky.  Em galaxies are plotted
as triangles; Ab galaxies as squares.}
\label{fig:surfbri}
\end{figure}

Figure~\ref{fig:surfbri} shows the mean and core surface brightness for
the 82 sample galaxies.  The core surface brightness is determined in the
most luminous 3$\times$3 pixel grid in the object, corresponding to
an area 1$\farcs$33 square on the sky; we compute the mean surface
brightness within the $R=23.5$ isophote.  We plot Em galaxies as triangles
and Ab galaxies as squares.  There is a clear trend of decreasing surface
brightness with increasing magnitude.  This trend makes the observation
of fainter objects more difficult, and leads to some undercounting of
faint sources, artificially depressing the faint end of the luminosity function.

The brightest galaxy (the cD N1129) is anomalous.
A cD galaxy is a giant elliptical with an extended low surface brightness 
envelope (Oemler 1976).  This envelope lowers the mean surface brightness
within the $R=23.5$ isophote for NGC~1129, since it occupies a large
fraction of the area within the limiting isophote.  The distended envelope
also makes the isophotal magnitude more sensitive to the sky subtraction
because the brightness profile approaches the sky level more gradually.

\subsection{Magnitude Segregation}

Magnitude (or equivalently luminosity) segregation is usually interpreted
as an indicator of mass segregation; the more massive (and hence more
luminous) galaxies are more centrally concentrated and move more slowly
than less massive (luminous) ones.  den Hartog \& Katgert (1996) find
luminosity segregation in 25 of their sample of 71 clusters, with a strong
signal in 10.

The central panel of Fig.~\ref{fig:sigmavsr2} shows $m_R$ as a function
of cluster radius in AWM~7, with a moving 11-galaxy average superimposed.
The median magnitude of galaxies within $r \lesssim 0.1h^{-1}$ Mpc is
brighter than the median outside this radius,
although the significance is low because of the small
sample size.  The radial extent of this luminosity excess is
roughly coincident (within a factor of 2) with NB's
value of $r_c = 51 \pm 3$ kpc for the X-ray core.  The extent of the
excess also matches the region of reduced velocity dispersion, suggesting
that the cold X-ray core, the reduced velocity dispersion, and the 
luminosity excess are related physical effects.

\subsection{Luminosity Function}

The most striking feature of the luminosity function (LF) of AWM~7
(Fig.~\ref{fig:sampmaghist}) is the peak near $R \sim 13.7$ and
the subsequent dip in galaxy counts near $m_R \sim 14.5$ ($M_R \sim -19.1$).
The cumulative distribution shows that although the peak is enhanced by
the binning, the dip is not an artifact.
A similar feature appears in the Coma cluster (Bernstein {\it et al.} 1995,
Biviano {\it et al.} 1995), in three of four moderate-redshift
Abell clusters studied by Wilson {\it et al.} (1997), and in a sample of
20 Abell clusters studied by Gaidos (1997).  Biviano {\it et al.} determine
Coma cluster membership spectroscopically (unlike Bernstein {\it et al.}, 
Wilson {\it et al.}, and Gaidos, who do so statistically), and suggest
that this feature may be common to rich clusters.

We attempt to characterize the cluster LF in terms of the Schechter (1976)
function parameters $\alpha$ (logarithmic faint-end slope) and $M_*$
(characteristic luminosity).
Measured values of $\alpha$ in clusters range from $-1.0$ (Lopez-Cruz
{\it et al.} 1997, Gaidos 1997) to $-2.2$ in $B$ and $I$ (De Propris
{\it et al.} 1995); the inclusion of dwarf galaxies
and low surface brightness galaxies increases the faint-end slope
(L\'opez-Cruz {\it et al.} 1997, Sprayberry {\it et al.} 1997).  In the
case of Coma, the inclusion of dwarf galaxies boosts the estimate of $\alpha$
from $-1.35$ to $-1.7$ (Trentham 1997).  Trentham (1997) argues that
since only dSph galaxies obey a
power-law distribution, the faint-end slope of the LF is a misleading
indicator, dominated primarily by its coupling to $M_*$.  He notes that
other galaxy types have bounded LFs, yet it is precisely these other
types that enter into most cluster LF determinations.

\begin{figure}[p]
\centerline{\epsfxsize=6.0in%
\epsffile{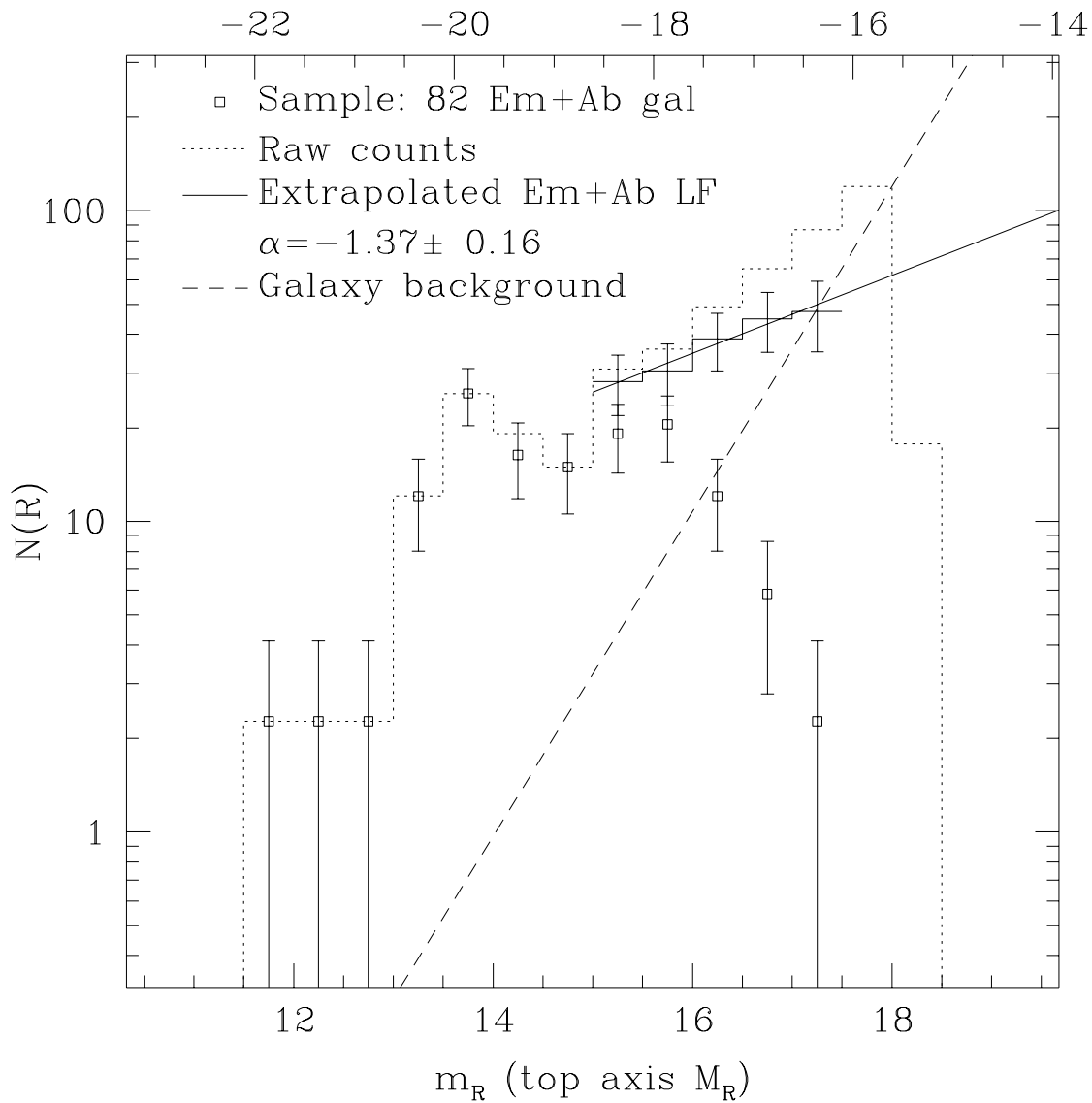}}
\caption{
 Raw galaxy counts (dotted histogram), sample galaxy counts (points with 
 error bars), $10^{0.6m}$ background (dashed line), and extrapolated
 LF (solid histogram and line).  The best power-law fit to the extrapolated
 LF corresponds to $\alpha=-1.36 \pm 0.16$ in the Schechter 
 parametrization.  The completeness limit is 16.0 .
 }
\label{fig:magalphaextrap}
\end{figure}

The Schechter function describes the LF of AWM~7 poorly; it cannot 
accommodate the peak and subsequent dip in the distribution.  
Maximum-likelihood fitting (Efstathiou \etal\ 1991) of a Schechter function
to the magnitude distribution (not shown)
forces $M_*$ to the peak near $R \sim 14$ and results in an ill-fitting
declining faint end, in contrast to the increasing counts seen in the last three
complete bins of the actual distribution.  Therefore, we obtain an estimate
of the faint-end slope of the LF by extrapolation.
We subtract a background field galaxy count, given by
$n(m) \propto 10^{0.6m}\, dm$ deg$^{-2}$ and normalized to the Century Survey,
from the observed galaxy counts (with or without measured redshifts)
to $R = 17.5$ (corrected for extinction), and fit a power law to the 
residual in the range $15.0 < R < 17.5$.  We plot the result in
Fig.~\ref{fig:magalphaextrap}.  The upper magnitude limit of the fit is
set by the incompleteness of the galaxy counts at faint magnitudes
due to poor star-galaxy separation on the CCD images with bad seeing.
The best-fit power law corresponds to a Schechter parameter 
$\alpha = -1.37 \pm 0.16$, with formal $\chi^2 / \nu = 3.57/6$.

We conclude that like Coma, AWM~7 has a LF with a bump at the bright end
and a steep faint end.  The steep faint end seen in field galaxy surveys
({\it e.g.} Marzke \etal\ 1994, Marzke \& da Costa 1997) is due to
blue galaxies.  In AWM~7, the LF is steep and red, populated by
absorption-line systems.  Mobasher \& Trentham (1998) find that the steep
($\alpha \sim -1.4$) $K$-band LF in Coma is due to dwarf spheroidals.
The issue is complicated
by surface-brightness selection effects: low-surface-brightness galaxies may
be missing from field surveys, which consequently underestimate
their contribution to the faint-end slope.

\subsection{Mass-to-Light Ratio}

Owing to the absence of precise photometry over a large area, the mass-to-light
ratio of AWM~7 has been poorly known.  Estimates have ranged from 160\,$h$ in
$V$ (in units of $M_\odot/L_\odot$) (Kriss {\it et al.} 1983, based on Einstein
IPC data) to 1120\,$h$ in $B$ (Beers {\it et al.} 1984, virial mass estimator).
Using ROSAT data, Dell'Antonio {\it et al.} (1995) find a value of 440\,$h$ in
$B$, and NB constrain the ratio to the range 400--1100\,$h$ in $B$.  Kriss
{\it et al.} (1983) find a range 140--200\,$h$ in $V$ for four other MKW and
AWM groups.  The main uncertainty in X-ray determinations of the mass is
introduced by the cooling flow.

We compute the mass from the virial estimator appropriate for the case of
galaxies embedded in a diffuse distribution of dark matter, with the added
assumption that the galaxies trace the dark matter distribution.
We exclude the Em galaxies
from the computations on the grounds that they constitute a dynamically
distinct population superposed on the virialized, Ab-populated cluster.
The appropriate estimator (Binney \& Tremaine 1987) is
$$ M_{\rm est} = {{3\pi N} \over {2G}}
                 {{\Sigma_{i=1}^N v_i^2 } \over
                  {\Sigma_{i=1}^N \Sigma_{j<i} |R_i-R_j|^{-1} } } $$
where $v_i$ is the radial velocity relative to the cluster mean, and $R_i$
is the projected distance from the cluster center.
This estimator assumes that the galaxies are in dynamical equilibrium within
the cluster potential, and that the galaxies trace the total mass.
If the dark matter is more extended than the galaxy distribution, this
prescription underestimates the $M/L$ ratio.  This mass estimate is
also very sensitive to the inclusion of foreground or background galaxies.
We estimate the error in the mass profile by the statistical ``jackknife''
procedure (Diaconis \& Efron 1983) as follows: within each projected radius,
we calculate the mass independently for all $n$ subsets of $n-1$ galaxies,
where $n$ is the total number of galaxies within said radius, and
with the velocities shuffled randomly for each subset.  We define the standard
deviation about the mean of the $n$ masses thus computed to be the error in
the mass estimate within that projected radius.

In principle this mass estimate should be adjusted by a surface term
(The \& White 1986) because the entire system is not included in the
observed sample.  Inclusion of this term requires 
knowledge of $\sigma(r)$, $N(r)$, and the dark matter profile.  The first
two factors can be constrained from the data, and the dark matter profile
can be reasonably described by model fits to hierarchical clustering simulations
as in Navarro \etal\ (1997).  However, our data set is not
extensive enough to support this analysis robustly; we do not have broad
enough angular coverage to self-consistently compute the core radius.
We have calculated the surface term, and find that the error is comparable
to the value of the correction itself.
Thus the masses we quote below do {\it not} incorporate a surface term.

\begin{figure}[p]
\centerline{\epsfxsize=6.0in%
\epsffile{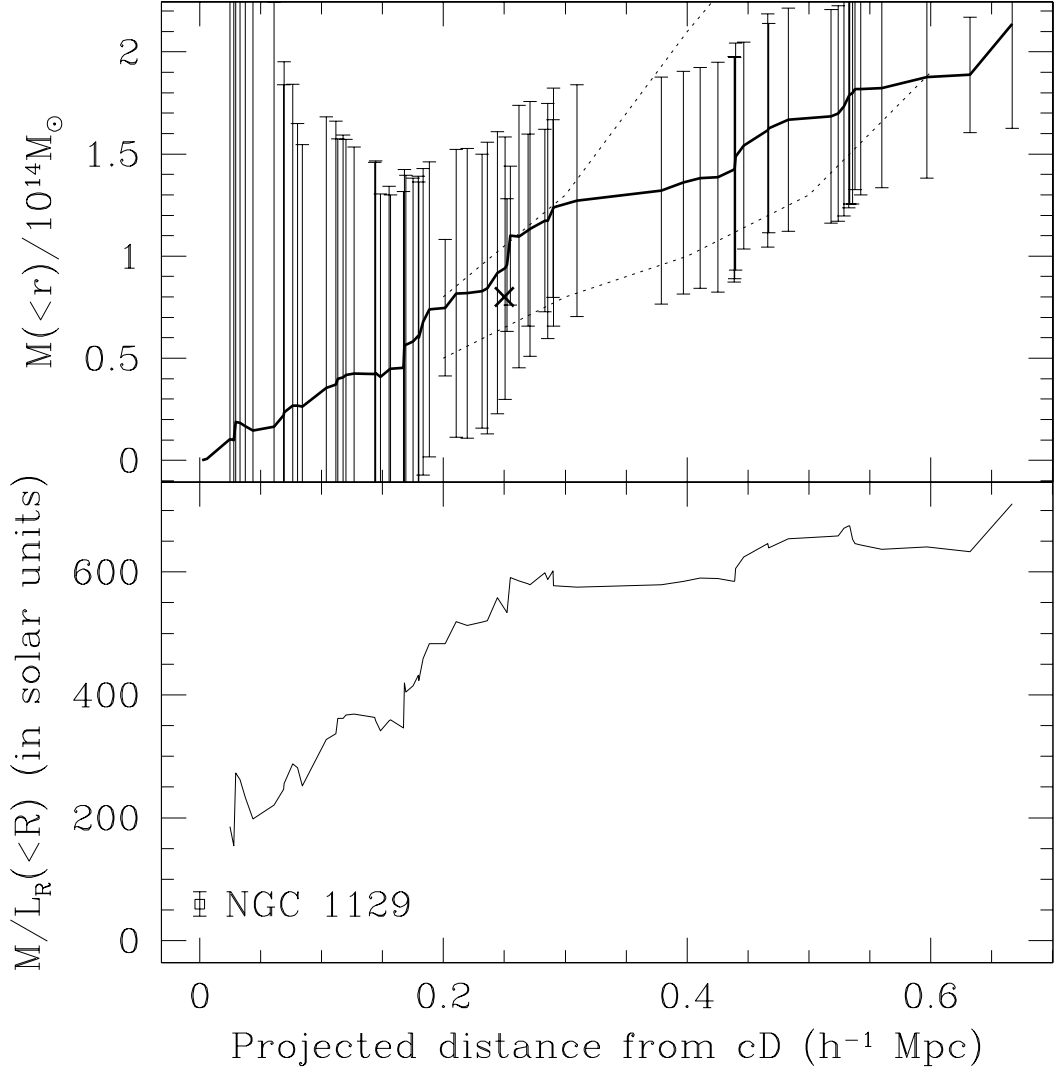}}
\caption{
 The enclosed virial mass (top) and $R$-band mass-to-light ratio (bottom) as
 a function of projected radius in the cluster.  The $\times$ marks the 
 data point from Dell'Antonio \etal\ (1995); the dotted lines indicate the
 profile of NB.  Errors on the mass are statistical jackknife estimates;
 they scale uniformly to errors on $M/L$.  The data point for NGC~1129 is
 from Bacon \etal\ (1985).  Note that $M/L_B$ = 1.58 $M/L_R$
 for Em galaxies with $B-R=1.5$, and that $M/L \propto h$.  
 }
\label{fig:masslight}
\end{figure}

The top panel of Figure \ref{fig:masslight} shows our integrated mass profile,
computed by applying the virial mass estimator to successively larger radii,
along with the profile determined by NB from a ROSAT temperature profile
and a data point from dell'Antonio \etal\ (1995).
The mass enclosed within 0.25 $h^{-1}$ Mpc is $\sim$9$\times 10^{13} M_\odot$, 
rising to $\sim$2$\times 10^{14} M_\odot$ within 0.6 $h^{-1}$ Mpc.
Our profile is in good agreement with the value derived by
Dell'Antonio {\it et al.} (1995) from X-ray data, 
who estimate $8 \times 10^{13}$ $h^{-1} \, M_\odot$ within 0.25 $h^{-1}$ Mpc;
it also lies within the errors of the NB profile for radii up to
$\sim$0.5$\,h^{-1}$ Mpc.  Beyond this radius, their mass estimates exceed ours;
Fig.~\ref{fig:masslight} shows their profile diverging from ours increasingly
at large radii.  They derive the mass profile beyond $\sim$600$\,h^{-1}$ Mpc by
extrapolation, however, and their luminosity, taken from Beers \etal\ 
(1984), is an underestimate which increases the computed $M/L$ ratio.

We compute the $R$-band light by adding up the luminosities of the
galaxies in our sample and correcting for incompleteness.  The correction
is required because
for any magnitude-limited sample, the observed luminosity is necessarily
an underestimate of the total cluster luminosity because the faintest
galaxies are not observed.  We correct for this incompleteness by
integrating the extrapolated luminosity function out to infinite magnitude.
For a Schechter function, the observed fraction of the total luminosity is
given by $\Gamma(\alpha+2,L_{\rm min}/L_*) / \Gamma(\alpha+2)$,
where $L_{\rm min}$ and $L_*$ are the luminosities corresponding to the 
completeness limit and $M_*$, respectively, and $\Gamma(x,y)$ is the
incomplete gamma function.
The completeness limit $R = 16.5$ is 3.6 mag fainter than
$M_* = -20.7$, yielding an observed luminosity fraction of $\sim$90\%.
A 0.5 mag error in the completeness limit corresponds to a 5\% error in
the observed luminosity fraction in this regime.
We observe $2.6 \times 10^{11}$ $L_\odot$ in the $R$-band within a projected
radius of 0.6 $h^{-1}$ Mpc, yielding a corrected total luminosity of
$2.9 \times 10^{11}$ $L_\odot$.  The cD alone contributes $\sim$14 \% of
the $R$-band luminosity within this radius.
 
We plot the $M/L$ profile in the lower panel of Fig.~\ref{fig:masslight}.
The data point for NGC~1129 is from Bacon \etal\ (1985), who tabulate
mass-to-light ratios for 197 ellipticals.  The mean $M/L_B$ in their
sample is 13; NGC~1129's ratio of $M/L_B = 94 \pm 31$ is the largest in
their sample.  We derive a $M/L_R$ ratio of
$\sim$600$\,h\,M_\odot/L_{\odot,R}$ for the cluster that remains
fairly constant outside a projected radius of 0.3~$h^{-1}$~Mpc, rising to
$\sim \! 650\pm170 \, h \, M_\odot/L_{\odot,R}$ near 0.45~$h^{-1}$~Mpc.
We include only the errors on the mass.  For comparison, note that
$M/L_B = 1.58 M/L_R$ because $(B-R)_\odot = 1.0$, but for elliptical
galaxies typically $B-R = 1.5$.  Our value of the mass-to-light ratio
is at the low end of the range of NB, where values are
based on extrapolations of X-ray temperature profiles to 1$^\circ$.  Within
0.25~$h^{-1}$ Mpc, Dell'Antonio {\it et al.} report 430$\,h$ in the $B$-band,
corresponding to 272$\,h$ in $R$; we find 530$\,h$.  Given the agreement in mass,
the discrepancy arises from differences in luminosity; Dell'Antonio \etal\
(1994) do not directly measure the luminosity in the cluster but instead
determine cluster membership by background subtraction, and then 
calibrate magnitudes derived from POSS plate scans against
Zwicky (1962) magnitudes, whose scatter is $\sim$0.3 mag (Bothun \& Cornell
1990, Geller \etal\ 1997).  Their largest source of error is the plate
photometry, particularly in light of the considerable variation of the
photographic sensitivity across the cluster, which introduces a large
systematic error in addition to the intrinsic scatter in the calibration
magnitudes.

The mass-to-light ratio of AWM~7 is lower in the center than at 
the periphery, because the depressed central velocity dispersion lowers the
mass estimate, and because there is excess luminosity in the core. 
It is interesting to note that
the $M/L$ profile at small radii approaches the value for the cD.
The profile also flattens outside $\sim$$0.3 \, h^{-1}$ Mpc, rising only
another 10\% out to $0.6 \, h^{-1}$ Mpc, suggesting that the dark matter
is less concentrated than the light.  The mass-to-light ratio levels off at
roughly twice the projected radius at which the velocity dispersion does.

Our calculated mass-to-light ratio for AWM~7 is in close agreement with
the values determined by Mohr {\it et al.} (1996b) for the clusters 
A~2626 and A~2440: they find $M/L_R \sim 610\,h$ and $660-880\,h$, respectively,
from a joint X-ray and optical study of the clusters.  Carlberg {\it et al.}
(1996, 1997) overlay 14 clusters to form an aggregate whose M/L ratio they
find to be $289 \pm 50\, h \, (M/L)_\odot$ in Gunn $r$.  Cirimele {\it et al.}
(1997) overlay 12 Abell clusters and find $M/L_V$ in the range 140--440 $h$.
Measurements of M/L ratios for distant clusters using
weak lensing yield a similar range of values:
Tyson \& Fischer (1995) find $M/L_V = 400 \pm 60 \, h \, (M/L_V)_\odot$ for
A~1689 at $z=0.18$, while Carlberg {\it et al.} (1994) report 225$\,h$ in the
$V$-band for a cluster at $z=0.325$, and 275$\,h$ for Coma, corrected for
``modest'' evolution of the galaxy LF.
In this context, AWM~7 no longer appears so exceptional.

\section{Discussion}

Our optical and spectroscopic survey of the central
$1.2 \times 1.2 \, h^{-1}$Mpc of AWM~7 yields 
a velocity dispersion profile, mass profile, luminosity function, and
mass-to-light profile of the cluster.  There is three-fold evidence for a cold
core in the cluster: the central velocity dispersion is depressed,
there is luminosity segregation on the same scale, with excess
luminosity in the core, and there is a cold X-ray core with similar scale.
The optically-determined mass is in
good agreement with X-ray determinations by NB and by Dell'Antonio \etal\
(1995).  Despite the offset in X-ray isophotes they see, we find no
kinematic evidence for substructure.  The luminosity function of AWM~7 is
peculiar: there is a dearth of galaxies with $R\sim14.5$, an excess of
galaxies just brighter, and a steeply rising faint end.  The faint end
is populated almost exclusively by red, absorption-line galaxies, in contrast
to the blue mIrr which dominate the steep faint end in the field Marzke \etal\
(1994).

\subsection{Evolution of the Luminosity Function and Formation of the cD}

The formation of cD galaxies is closely tied to
questions of LF evolution, universality, and mass segregation.  Theories
for cD formation include merging of dwarf galaxies through 
dynamical friction (Ostriker \& Tremaine 1975; White 1976; Ostriker \&
Hausmann 1977), cannibalization of neighboring galaxies (Gallagher \& Ostriker
1972; Richstone 1975, 1976), primordial origin (Merritt 1984), and mergers of
large, bright galaxies early in the cluster history, with additional
growth from accumulation of tidal debris or from cooling flows (Fabian \&
Nulsen 1977; Cowie \& Binney 1977; Fabian \etal\ 1984).

Active merging in a cluster would result in substantial evolution of the LF.
Thus the LF could in principle be used as an estimator of the dynamical age
of the cluster.  Although the traditional ``cannibalization'' merger scenarios
held that dwarf galaxies agglomerate into larger ellipticals (see Barnes \&
Hernquist 1992a for a review), recent simulations (Barnes 1992, Dubinski 1997)
indicate that larger galaxies tend to merge, with an abundance
of faint galaxies condensing from the tidal tails produced in
these interactions (Barnes \& Hernquist 1992b).  This scenario can
explain many features of the LF: the dip may reflect depletion of galaxies
through merging; the peak brightward of the dip has a dynamical
origin --- it is populated
by products of these mergers; and the steep faint end arises from the dwarfs
formed from the tidal tails of the mergers.

Early theories that mergers deplete the dwarf population should result in 
a paucity of red dwarfs, particularly in clusters with low velocity
dispersion where mergers are
more efficient (David \& Blumenthal 1992).  High-dispersion clusters may
have lower merger rates and thus more red dwarfs at the current epoch.
In this paradigm, AWM~7's high velocity dispersion would suppress merging and
would thus account for the steep, red faint end, but could not explain
the other features of the LF, including the bump and the presence of the cD.

AWM~7 presents a steep, red dwarf population in a high-dispersion cluster,
which argues against the merger of pre-existing dwarfs as the source of the
cD, both because (1) mergers are {\it a priori} unlikely due to the high
$\sigma$, and (2) mergers of dwarfs to form the cD would deplete the faint end.
Unless AWM~7 initially had an even steeper faint end, this scenario seems
unlikely.

\section{Conclusion}

Our study of AWM 7 reveals two important features: the cluster has a cold
core, and the steeply rising faint end of the LF is populated predominantly by
absorption-line galaxies, in contrast to the emission-line galaxies that
populate
the faint end of the field LF.  In AWM~7, the emission galaxies are probably
a dynamically distinct infalling population superposed on the relaxed
system of absorption-line galaxies; the little substructure that is
apparent in the velocity data is entirely attributable to the emission
galaxies. We have resolved the anomalous earlier mass-to-light ratio 
calculated for AWM~7; our value
($\sim \! 650\pm170 \, h \, M_\odot/L_{\odot,R}$ at 0.45~$h^{-1}$~Mpc)
is concordant with those of similar systems.  The mass-to-light ratio
approaches the central cD's value at small radii, and is flat at large
radii.

The proximity of AWM~7 allows for
direct determination of the LF (with redshifts rather than by statistical
background subtraction) well below $L_*$, since $L_*$ corresponds
to $R \sim$ 13 at 5000 \kms.  A deeper survey
will result in denser sampling in the core of the cluster, aiding in
discriminating between dynamical models, and will boost the signal in
the various tests we have performed here.  Deeper surveys will also
yield direct measurements of the faint end slope of the LF.


\acknowledgments

We thank Scott Kenyon and Michael Kurtz for their assistance, and Susan Tokarz
for reducing the FAST spectroscopic data.
This work is supported by the Smithsonian Institution.
DMK was supported by a National Science Foundation Graduate Fellowship.


\clearpage

\begin{table}
\dummytable \label{tab:datasample}
\end{table}

\setcounter{page}{26}

\clearpage

\clearpage

\clearpage

\clearpage
 
\clearpage

\clearpage

\clearpage

\clearpage

\clearpage

\clearpage

\clearpage

\clearpage

\end{document}